\documentstyle[aps,epsf,floats]{revtex}

\begin{document}

%%%
\twocolumn[{
%%%
\widetext

\title{%%
Phase diagram of the one-dimensional %%
Holstein model of spinless fermions%%
}

\author{%%
Robert J. Bursill$^{*}$, Ross H. McKenzie and Chris J. Hamer%%
}

\address{
School of Physics, University of New South Wales, Sydney, NSW 2052, %%
Australia.%%
}

\maketitle

%%%
\mediumtext

\begin{abstract}

The one-dimensional Holstein model of spinless fermions interacting with 
dispersionless phonons is studied using a new variant of the density 
matrix renormalization group. By examining various low-energy 
excitations of finite chains, the metal-insulator phase boundary is 
determined precisely and agrees with the predictions of strong 
coupling theory in the anti-adiabatic regime and is consistent with 
renormalization group arguments in the adiabatic regime. The Luttinger 
liquid parameters, determined by finite-size scaling, are consistent 
with a Kosterlitz-Thouless transition.

\end{abstract}

%71.45.Lr Charge-density-wave systems
%71.38.+i Polarons and electron-phonon interactions (see also 63.20.K)
%63.20.Kr Phonon-electron interactions
%71.30.+h Metal-insulator transitions
%%

%%%
}] \narrowtext

The challenge of understanding superconductivity in fullerenes, bismuth 
oxides, and the high-$T_{\text c}$ cuprates has renewed interest in 
models of interacting electrons and phonons \cite{Mott}. Unlike
conventional metals these materials are not necessarily in the
weak-coupling regime where perturbation theory can be used or the 
strong-coupling regime in which a polaronic treatment is possible 
\cite{Mott}. Neither are they necessarily in the adiabatic regime in 
which characteristic phonon energies are much less than characteristic 
electronic energies. This challenge has led to numerical studies of the 
Holstein (or molecular crystal) model of electrons  interacting with 
dispersionless phonons %\cite{Holstein}
in infinite dimensions, %\cite{Freericks}
two dimensions, %\cite{Gubernatis}
one dimension %\cite{Marsiglio}
and on just two sites %\cite{Alexandrov}
(see the references in \cite{Mott,McKenzie3}). The 
one-dimensional case is important because of the wide range of
quasi-one-dimensional materials which undergo a Peierls or
charge-density-wave (CDW) instability due to the electron-phonon 
interaction.
%An example which is currently receiving considerable 
%attention is the spin-Peierls compound CuGeO$_3$ \cite{Hase}.
Most theoretical treatments assume the adiabatic limit 
and treat the phonons in a mean-field approximation. However, it has 
been argued that in many CDW materials the quantum lattice fluctuations 
are important \cite{McKenzie1}.

In this Letter we present a study of the one-dimensional Holstein model 
of spinless fermions at half-filling using the density matrix 
renormalization group (DMRG). 
%\cite{White,Caron,Jeckelmann1,Jeckelmann2}.
This model is 
particularly interesting because at a finite fermion-phonon coupling 
there is a quantum phase transition from a Luttinger liquid (metallic) 
phase to an insulating phase with CDW long-range order 
\cite{Hirsch_Fradkin,Lebowitz}. This illustrates how quantum 
fluctuations can destroy the Peierls state. The Hamiltonian is
\begin{eqnarray}
{\cal H} & = &
- t \sum_{i=1}^{N} \left( c_i^\dagger c_{i+1} + c_{i+1}^\dagger c_i 
\right)
+ \omega \sum_{i=1}^{N} a_i^\dagger a_i
%
%%%
\nonumber
%%%
\\
%
%%%
& &
-
%%%
\;
g \sum_{i=1}^{N} \left(c_i^\dagger c_i - \frac{1}{2} \right)
 \left(a_i + a_i^\dagger \right),
\label{H}
\end{eqnarray}
where $c_i$ destroys a fermion on site $i$, $a_i$ destroys a local 
phonon of frequency $\omega$, $t$ is the hopping integral, $g$ is the 
fermion-phonon coupling and a periodic chain of $N$ sites is assumed. 
The phase transition occurs at a critical coupling $g_{\text c}$ 
separating metallic ($0 \leq g \leq g_{\text c}$) and CDW insulating 
phases
($g > g_{\text c}$) \cite{Hirsch_Fradkin,Lebowitz}. In the strong 
coupling limit ($g^2 >> \omega t$) (\ref{H}) can be mapped onto the 
anisotropic, antiferromagnetic Heisenberg ($XXZ$) model 
\cite{Hirsch_Fradkin} which is exactly soluble. The transition occurs at 
the spin isotropy point, is of the Kosterlitz-Thouless (K-T) type, and 
the Luttinger liquid parameters can be found in the metallic phase 
\cite{McKenzie3}.

%The model has been studied numerically by world line (WLMC) 
%\cite{Hirsch_Fradkin} and Green's function (GFMC) \cite{McKenzie3} 
%Monte Carlo methods. As can be seen from Fig.\ \ref{phase_diagram}, the 
%MC results for $g_{\text c}$ are reasonably consistent with one another 
%and with strong coupling theory for small $t/\omega$. For large 
%$t/\omega$, however, the two MC calculations give markedly different 
%phase boundaries. Also, in \cite{McKenzie3} there was some evidence 
%that the transition is not of the K-T type in this region.

%%%%%%%%%%%%%%%%%%FIGURE 1%%%%%%%%%%%%%%%%%%%%%%%%%%
\begin{figure}[htbp]
%
%%%
\centerline{\epsfxsize=8.4cm \epsfbox{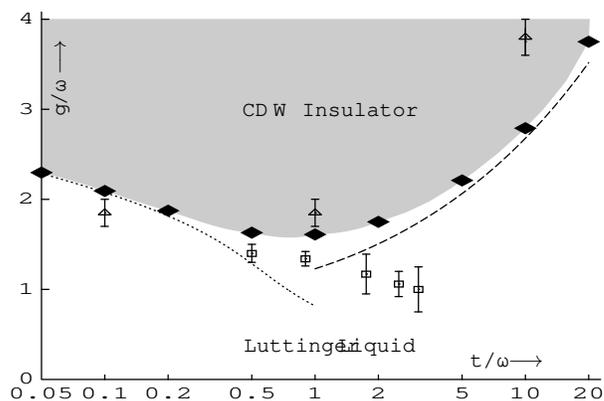}}
%\epsfbox{fig1.ill}
%
\caption{
Zero temperature phase diagram of the one-dimensional Holstein model of 
spinless fermions at half-filling. For small fermionic-phonon coupling 
$g$ the system is a Luttinger liquid with parameters that vary with the 
coupling. For large $g$ the system has an energy gap and long-range 
charge-density-wave (CDW) order. The solid diamonds denote the phase 
boundary from this DMRG study. The systematic errors are smaller than 
the diamonds. The results of previous quantum Monte Carlo studies are 
denoted by squares \protect\cite{Hirsch_Fradkin} and triangles 
\protect\cite{McKenzie3}. The dotted line is the phase boundary from 
strong coupling theory \protect\cite{Hirsch_Fradkin} and the dashed line 
is defined by
$\omega = \Delta_{\text{MF}}$
(where
$\Delta_{\text{MF}} \equiv 8t e^{-\pi t \omega / g^2}$
is the mean-field energy gap) and is the approximate location of the 
phase boundary predicted by a two-cutoff renormalization group scheme 
\protect\cite{Caron2}.
}
\label{phase_diagram}
\end{figure}

The phase diagram of (\ref{H}) over a wide range of adiabacity 
parameters ($0.05 \leq t/\omega \leq 20$) is shown in Fig.\ 
\ref{phase_diagram}. A new variant \cite{Bursill} of the DMRG method 
\cite{White,Caron,Jeckelmann1,Jeckelmann2} is used to determine 
the energy of low-lying excitations to a far greater precision than 
previous quantum Monte Carlo (QMC) studies 
\cite{McKenzie3,Hirsch_Fradkin}. Finite-size scaling (FSS) of a number 
of energy gaps permits the accurate determination of $g_{\text c}$ and 
the 
Luttinger liquid parameters.
%close to the transition point.

The study of fermion-boson models such as (\ref{H}) by exact 
diagonalization or the DMRG presents a challenge because there are an 
infinite number of phonon quantum states on each site. Caron and 
Moukouri have studied the $XY$ spin-Peierls and
free acoustic phonons models \cite{Caron} on open chains using a 
conventional DMRG algorithm. The simple truncation of the phonon Hilbert 
space used in these calculations can require an excessively large number 
of states, to the extent where the effort expended in representing a 
single site becomes comparable to that expended in representing a block. 
This becomes important when trying to study periodic systems (which are 
more useful for FSS studies) where an extra site is usually added to 
avoid direct interactions between blocks. Jeckelmann and White devised a 
scheme that maps bosons onto fermions which they applied to the polaron 
problem (a single electron interacting with the phonons) in one and two 
dimensions \cite{Jeckelmann1}. A more promising method, which 
dramatically reduces the number of states required to represent a site, 
has been used to examine small (6 site), half-filled Holstein systems 
using exact diagonalization \cite{Jeckelmann2}. We have developed a 
somewhat similar DMRG algorithm which is designed to solve
{\em periodic} systems with a large number of degrees of freedom per 
site. The details of the method will be published elsewhere 
\cite{Bursill}---here we concentrate on the results for (\ref{H}).

The good quantum numbers used are the total fermion number
$ \hat{N} \equiv \sum_{i=1}^N c_i^{\dagger} c_i $,
and, for the neutral case ($\frac{1}{2}$-filled band; $\hat{N} = N/2$), 
the parity (particle-hole) operator
$\hat{P}: c_i \mapsto (-1)^i c_i^{\dagger}; a_i \mapsto -a_i$
The energies calculated are the ground state energy
$E_{\text G} \equiv E_0(\hat{N} = N/2, \hat{P} = 1)$,
the charge gap 
$\Delta_{\text{ch}} \equiv E_0(\hat{N} = N/2 \pm 1) - E_{\text G}$,
and the 1- and 2- photon gaps (the two lowest neutral excitations 
\cite{notation})
$\Delta_1 \equiv E_0(\hat{N} = N/2, \hat{P} = -1) - E_{\text G}$
and 
$\Delta_2 \equiv E_1(\hat{N} = N/2, \hat{P} = 1) - E_{\text G}$.
A number of accuracy checks were performed: The DMRG reproduces exact 
results in the non-interacting and strong coupling limits, and the DMRG 
results agree with QMC results for systems of up to $N = 16$ sites 
\cite{McKenzie3} within error bars. The DMRG accuracy is determined 
by the parameter $m$---the number of density matrix eigenstates retained 
per block. Table \ref{convergence} lists convergence results for 
$\Delta_{\text{ch}}$, along with the QMC results \cite{McKenzie3}. The 
DMRG errors, being systematic rather than statistical, are two to three 
orders of magnitude smaller than the QMC errors.
%Similar accuracy is observed for the other energies calculated.

%%%%%%%%%%% TABLE 1 %%%%%%%%%%%%%%
\begin{table}[htbp]
%
%%\vspace{0.2cm}
%
%%\begin{center}
%
\caption{Convergence of the charge gap $\Delta_{\text{ch}}$ with the 
DMRG 
truncation parameter $m$ for various system sizes $N$ using parameters 
$t=\omega$ and $g=1.5\omega$. QMC results \protect\cite{McKenzie3} are 
included
for comparison.}
\begin{tabular}{c|cccc}
%
%%\hline
%%\hline
%
$m$ & $N=4$ & $N=8$ & $N=16$ & $N=32$ \\ 
\hline
26   &   0.4110  &   0.1971   &   0.1021   &   0.05504 \\
36   &   0.4110  &   0.1971   &   0.1004   &   0.05244 \\
48   &   0.4110  &   0.1971   &   0.1002   &   0.05148 \\
66   &   0.4110  &   0.1971   &   0.1002   &   0.05117 \\
78   &   0.4110  &   0.1971   &   0.1001   &   0.05103 \\
94   &   0.4110  &   0.1971   &   0.1001   &   0.05099 \\
\hline
QMC & 0.416(4) & 0.200(9) & 0.06(3) &      --   \\
%
%%\hline
%%\hline
%
\end{tabular}
%
%%\end{center}
%
\label{convergence}
\end{table}

\begin{figure}[htbp]
%
%%%
\centerline{\epsfxsize=8.4cm \epsfbox{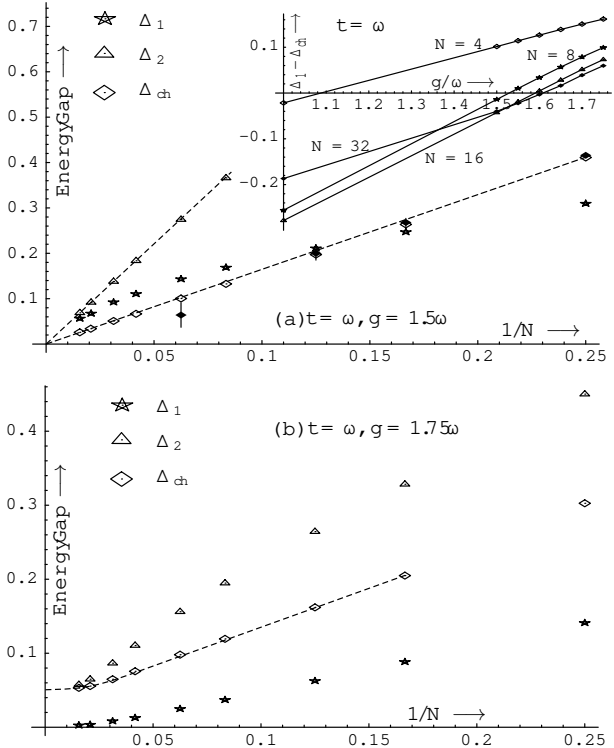}}
%\epsfbox{fig2.ill}
%
\caption{
Finite-size scaling of the different energy gaps in the (a) Luttinger 
liquid and (b) insulating phases. The charge gap ($\Delta_{\text{ch}}$) 
and 
one- and two- photon gaps ($\Delta_1$ and $\Delta_2$) are plotted as 
functions of the inverse lattice size $1/N$ for $t=\omega$ and (a) 
$g=1.5\omega$ and (b) $g=1.75\omega$. Also shown in case (a) is 
$\Delta_{\text{ch}}$ as calculated using QMC \protect\cite{McKenzie3} 
(solid 
diamonds with error bars). In (a) the dashed lines are straight lines 
through the origin, the slope of which can be used to extract the 
Luttinger liquid exponent $K_\rho$ (see Fig.\ 
\protect\ref{Luttinger_parameters}). In (b) the dashed line is a guide 
for the 
eye. The differences between the two phases are seen in: (i) The 
relative size of $\Delta_{\text{ch}}$ and $\Delta_1$ is opposite for the 
two phases. (ii) In the limit $N \rightarrow \infty$ the gaps 
extrapolate to zero (a non-zero value) in the Luttinger liquid 
(insulating) phase. The inset to (a) shows the difference
$\Delta_1 - \Delta_{\text{ch}}$
as a function of the coupling for various $N$. The critical coupling 
$g_{\text c}$ is determined as the value at which this difference 
vanishes 
in the limit $N \rightarrow \infty$.
}
\label{gaps}
\end{figure}

Typical FSS plots of the various energy gaps are shown in Fig.\ 
\ref{gaps} for the metallic $(g < g_{\text c})$ and insulating
$(g > g_{\text c})$ phases. In the metallic phase the gaps vanish 
linearly 
with $1/N$ as $N \rightarrow \infty$, with $\Delta_1$ lying {\em above} 
$\Delta_{\text{ch}}$ for large $N$. In the insulating phase 
$\Delta_1$ lies {\em below} $\Delta_{\text{ch}}$ and 
$\Delta_{\text{ch}}$ and 
$\Delta_2$ approach non-zero values as $N \rightarrow \infty$ whilst 
$\Delta_1$ rapidly tends to zero, the state
$E_0(\hat{N} = N/2, \hat{P} = -1)$
being asymptotically degenerate with the ground state in this phase.
%\cite{asymptotic}.

%%%%%%%%%%%%FIGURE 2%%%%%%%%%%%%%%%%%%%%%%%%%%%%%%

In the QMC studies \cite{McKenzie3,Hirsch_Fradkin} the critical point 
$g_{\text c}$ was determined as the point at which an order parameter or 
the charge gap $\Delta_{\text{ch}}$ becomes non-zero. However, in a K-T 
transition these quantities behave as
$\Delta_{\text{ch}} \sim e^{ -A(g - g_{\text c})^{-1} }$,
and there are nonlinear corrections to FSS which make the precise 
determination of $g_{\text c}$ very difficult by this method. Our method 
of determining $g_{\text c}$ is inspired by work on the frustrated 
Heisenberg model \cite{Affleck} where the transition point was 
determined by the crossover of singlet and triplet gaps. It is known 
that K-T transitions have a hidden $SU(2)$ symmetry \cite{Nomura}. We 
hypothesise that
%the model (\ref{H}) becomes spin rotationally invariant 
at $g = g_{\text c}$, the states
$E_0(\hat{N} = N/2 \pm 1)$
and
$E_0(\hat{N} = N/2, \hat{P} = -1)$
form a degenerate triplet in the thermodynamic limit. Plots of the 
difference
$\Delta_1 - \Delta_{\text{ch}}$
are included in the inset of Fig.\ \ref{gaps} for various $N$. A 
crossover point $g_{\text c}(N)$ is defined as the $g$ value at which
$\Delta_1 = \Delta_{\text{ch}}$. 
$g_{\text c}(N)$, listed in Table \ref{criticality} for various values 
of 
$t/\omega$, approaches $g_{\text c}$ as $N \rightarrow \infty$ 
\cite{differences}.
%The convergence is rapid enough to permit a reasonably precise
%determination of $g_{\text c}$.
The combined errors (DMRG truncation, discretization and fitting in
$g$, and extrapolation to $N = \infty$) are estimated to be
less than five precent.

%%%%%%%%%%% TABLE 3 %%%%%%%%%%%%%%
\begin{table}[htbp]
\caption{Convergence of the crossover point $g_{\text c}(N)/\omega$, 
determined by $\Delta_{\text{ch}}=\Delta_1$, with the system size $N$ 
for 
various hopping parameters $t$.}
\begin{center}
\begin{tabular}{c|ccccccc}
%
%%\hline
%%\hline
$N$               &    4     &    8     &    16    &    32    &   64   & 
128     &  256  \\
\hline
$t = 0.1 \omega$  &  2.0878  &  2.0911  &  2.0920  &          &        & 
        &       \\
$t = \omega$      &  1.087   &  1.528   &  1.591   &   1.608  &  1.613 & 
       &        \\
$t = 10 \omega$   &          &          &          &   2.220  &  2.649 & 
2.765  & 2.788  \\
%%\hline
%%\hline
%
\end{tabular}
\end{center}
\label{criticality}
\end{table}

The resulting phase boundary is shown in Fig.\ \ref{phase_diagram}, 
along with the two QMC calculations \cite{McKenzie3,Hirsch_Fradkin}, and 
the result of strong coupling theory \cite{Hirsch_Fradkin} which becomes 
exact as $t \rightarrow 0$. The DMRG results agree well with the strong 
coupling curve for $t/\omega < 0.2$. For large $t$ the results lie close 
to the curve defined by
$\omega = \Delta_{\text{MF}} \equiv 8t e^{-\pi t \omega / g^2}$.
This curve was predicted to be the approximate phase boundary for
$t > \omega$ within a two-cutoff renormalization scheme, where
$\Delta_{\text{MF}}$ is the mean-field energy gap \cite{Caron2}. A 
saddle-point 
expansion about the mean-field solution \cite{Wu} suggests that there is 
a first-order transition for $\omega \sim \Delta_{\text{MF}}$.

\begin{figure}[htbp]
%
%%%
\centerline{\epsfxsize=8.4cm \epsfbox{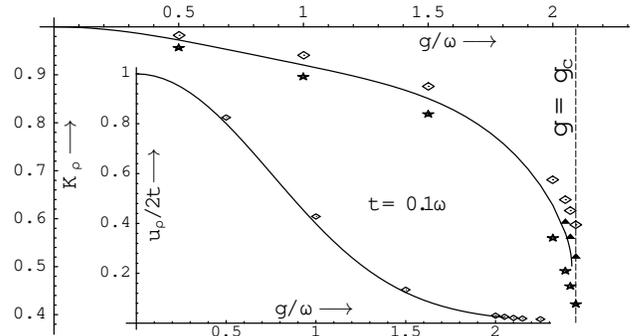}}
%\epsfbox{fig3.ill}
%
\caption{
Coupling dependence of the Luttinger liquid parameters $u_\rho$ (charge 
velocity) and $K_\rho$ (correlation exponent) for the case 
$t=0.1\omega$. The diamonds and stars are the values of $K_\rho$ 
calculated from the finite-size scaling of the energy gaps 
$\Delta_{\text{ch}}$ and $\Delta_1$, respectively. The solid triangles 
are 
the $K_\rho$ values determined from (\protect\ref{combined}). The solid 
curves 
are the results of strong coupling theory \protect\cite{McKenzie3}.
}
\label{Luttinger_parameters}
\end{figure}

We next investigate the nature of the transition and the Luttinger 
liquid parameters in the metallic phase $(0 \leq g \leq g_{\text c})$. 
For 
a Luttinger liquid of spinless fermions, $E_{\text G}$ scales according 
to
$\frac{E_{\text G}}{N} \sim \epsilon_\infty - \frac{\pi u_\rho}{6 N^2}$
\cite{Voit}, where $\epsilon_\infty$ is the bulk ground state energy 
density and $u_\rho$ is the charge velocity. From conformal field theory 
\cite{Cardy} the scaling forms for the gaps are:
$\Delta_{\text{ch}} \sim \frac{\pi u_\rho}{2 K_\rho N}$
and
$\Delta_1, \Delta_2 \sim \frac{2 \pi u_\rho K_\rho}{N}$,
where $K_\rho$ is the correlation exponent. The crossover method of 
determining $g_{\text c}$ is equivalent to the assumption that
$K_\rho = \frac{1}{2}$ at $g = g_{\text c}$, i.e., the transition is of 
the K-T type \cite{Shankar}. In Fig.\ \ref{Luttinger_parameters} 
$u_\rho$ (determined from the FSS of $E_{\text G}$) and $K_\rho$ (the 
values determined from the FSS of both $\Delta_{\text{ch}}$ and 
$\Delta_1$) are shown as functions of $g/\omega$ for the case
$t=0.1\omega$. The $u_\rho$ values agree very well with strong coupling 
theory. The agreement for $K_\rho$ is not as good, due to the presence 
of nonlinear correction terms to the energy gap scaling forms.

%%%%%%%%%%%%%%%%%%%%%%FIGURE 3%%%%%%%%%%%%%%%%%%%%%

A theory for these nonlinear correction terms has been developed for 
the critical case \cite{Affleck,Cardy1}, namely
$\Delta_{\text{ch}} \sim \frac{2 \pi u_\rho}{N}
\left[ \frac{1}{4 K_\rho} + \frac{A}{ \log N } + \ldots 
\right]$
and
$\Delta_2 \sim \frac{2 \pi u_\rho}{N}
\left[ K_\rho - \frac{3A}{ \log N } + \ldots \right]$,
where $A$ is a constant and $K_\rho = 1/2$. By taking 
the combination
\begin{equation}
3 \Delta_{\text{ch}} + \Delta_2 \sim \frac{2 \pi u_\rho}{N}
\left[ \frac{3}{4 K_\rho} + K_\rho  + \ldots \right],
\label{combined}
\end{equation}
the leading nonlinear correction is cancelled at $g=g_{\text c}$, the 
next correction being $O \left( \frac{1}{(\log N)^2} \right)$. For
$t = \omega$ and $g = g_{\text c}$ $K_\rho = 0.52$ is obtained if 
(\ref{combined}) is used to determine $K_\rho$. In comparison, values of 
0.59 and 0.42 are obtained from the scaling of $\Delta_{\text{ch}}$ and 
$\Delta_1$, respectively. 
It might be expected that (\ref{combined}) should give better results 
for 
$K_\rho$ around the critical point than the scaling of 
$\Delta_{\text{ch}}$ 
or $\Delta_1$. The resulting values, plotted in Fig.\ 
\ref{Luttinger_parameters}, are in good agreement with strong coupling 
theory. To check the consistency of the transition with a K-T transition 
the value of $g$ at which $K_\rho$ (calculated using (\ref{combined}))
equals $\frac{1}{2}$ is listed in Table \ref{consistency}. It can be
seen that the transition is consistent with a K-T transition throughout
the phase diagram.

%%%%%%%%%%% TABLE 3 %%%%%%%%%%%%%%
\begin{table}[htbp]
\caption{Transition point $g_{\text c}$ (as determined by the crossover 
of 
$\Delta_{\text{ch}}$ and $\Delta_1$) and $g^*$, the value of $g$ at 
which
$K_\rho = \frac{1}{2}$, (where $K_\rho$ is calculated from 
(\protect\ref{combined})) for various hopping parameters $t$. The 
agreement 
between $g_{\text c}$ and $g^*$ is consistent with the transition being 
of 
the K-T type.}
\begin{center}
%
%\begin{tabular}{c|cccccccc}
\begin{tabular}{c|cccccc}
$t/\omega$ &
0.05       &      0.1     &     0.5      &       1      & 
5      &      10     \\
\hline
$g_{\text c}/\omega$ &
2.297(2)   &   2.093(2)   &   1.63(1)    &   1.61(1)    & 
2.21(3)    &   2.79(5)   \\
$g^* / \omega$ &
2.299      &    2.102     &     1.64     &     1.62     & 
2.27     &      2.89   \\
%
%%\hline
%%\hline
%
\end{tabular}
\end{center}
\label{consistency}
\end{table}

\begin{figure}[htbp]
%
%%%
\centerline{\epsfxsize=8.4cm \epsfbox{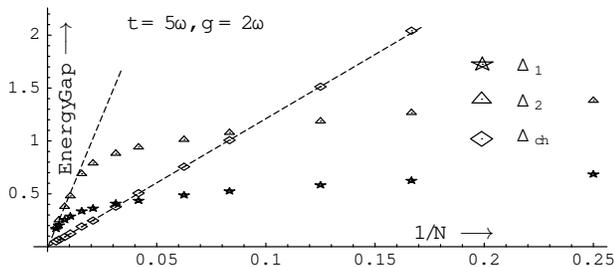}}
%\epsfbox{fig4.ill}
%
\caption{
The energy gaps $\Delta_{\text{ch}}$, $\Delta_1$ and $\Delta_2$ as 
functions of $1/N$ for $t=5\omega$ in a metallic case $g=2\omega$. 
Dashed lines are straight lines through the origin. The gaps $\Delta_1$ 
and $\Delta_2$ are not linear in $1/N$ because in the adiabatic regime 
there is strong mixing between fermionic and phonon excitations. For 
systems of less than 20 sites the lowest excitations with quantum 
numbers $\hat{P} = \pm 1$ and $\hat{N} = N/2$ are predominantly 1-
and 2- phonon excitations.
}
\label{softening}
\end{figure}

Finally, we consider the question of phonon softening and the mixing of 
phonon and fermion excitations. Fig.\ \ref{softening} shows the FSS of 
the energy gaps for a metallic case ($g < g_{\text c}$) with large 
hopping $t = 5 \omega$. Whilst $\Delta_{\text{ch}}$ is linear in $1/N$, 
$\Delta_1$ and $\Delta_2$ are highly nonlinear. This is because the 
lowest fermionic and bosonic, neutral excitations have the same quantum 
numbers, those of $\Delta_1$ and $\Delta_2$: The non-interacting 
fermionic gap $\frac{4 \pi t}{N}$ only becomes less than the bare phonon 
frequency $\omega$ for
$N \approx \frac{4 \pi t}{\omega} \approx 60$
and thus $\Delta_1$ and $\Delta_2$ are predominantly 1- and 2- phonon 
excitations for small $N$ (flat in $1/N$), only becoming 1- and 2- 
particle-hole excitations (linear in $1/N$) for large $N$. Note that 
for these parameter values the phonons are softened---the renormalised 
phonon frequency is around half the bare phonon frequency $\omega$. It 
would be interesting to calculate the 1-phonon Green's function to see 
if the phonons soften completely at the transition. The 2-phonon Green's 
function could be used to study phonon anharmonicity.

%%%%%%%%%%%%%FIGURE 4%%%%%%%%%%%%%%%%%%

In conclusion, we have shown that, using a new variant of the DMRG, 
the phase boundary of the one-dimensional Holstein model of spinless
fermions can be accurately determined. The transition is consistent with
a K-T transition over a wide range of adiabacity. In the antiadiabatic 
limit the phase boundary and Luttinger liquid parameters agree well with 
strong coupling theory. In the adiabatic limit the phase boundary lies 
close to a curve predicted by renormalization group arguments. 
Challenges that remain include: 1) finding a method of cancelling
nonlinear corrections to scaling, and hence accurately calculating the 
correlation exponent $K_\rho$, in the whole of the Luttinger liquid 
regime; 2) developing a theory of FSS when the conformally invariant 
field is coupled to a dispersionless field with a gap in order to 
explain the nonlinear scaling in Fig.\ \ref{softening}; and 3) a 
detailed investigation of phonon softening and anharmonicity.
%We hope to address some of these questions in future studies and to
%apply our numerical techniques to other electron-phonon models.

This work was supported by the Australian Research Council. We 
thank J. Voit, H. Eckle, E. Jeckelmann, T. Xiang, H. Fehske and
V. Kotov for useful discussions. Calculations were performed at
the New South Wales Centre for Parallel Computing.

%%\begin{thebibliography}{99}

%%


\begin{references}

\bibitem[*]{email}
Email address: ph1rb@newt.phys.unsw.edu.au

\bibitem{Mott}
A. S. Alexandrov and N. Mott, Polarons and Bipolarons (World Scientific, 
Singapore, 1995).

%\bibitem{Holstein}
%T. Holstein, Ann.\ Phys.\ {\bf 8}, 325, 343  (1959).

%\bibitem{Freericks}
%J. R. Freericks, M. Jarrell, and D. J. Scalapino, Phys.\ Rev.\ B
%{\bf 48}, 6302 (1993), and references therein.

%\bibitem{McKenzie2}
%K. Kim, R.H. McKenzie and J.W. Wilkins, Phys.\ Rev.\ Lett.\
%{\bf 71}, 4015 (1993).

%\bibitem{Gubernatis}
%P. Niyaz, J. E. Gubernatis, R. T. Scalettar, and C. Y. Fong, Phys.\ 
%Rev.\ B {\bf 48}, 16011 (1993), and references therein.

%\bibitem{Marsiglio} F. Marsiglio, Physica C {\bf 244}, 21 (1995).

%\bibitem{Alexandrov}
%A. S. Alexandrov, V. V. Kabanov, and D. K. Ray, Phys.\ Rev.\ B
%{\bf 49}, 9915 (1994), and references therein.

\bibitem{McKenzie3}
R.H. McKenzie, C.J. Hamer and D.W. Murray, Phys.\ Rev.\ B {\bf 53}, 9676 
(1996).

%\bibitem{Hase}
%M. Hase et.\ al., Phys.\ Rev.\ Lett.\ {\bf 70}, 3651 (1993).

\bibitem{McKenzie1}
R.H. McKenzie and J.W. Wilkins, Phys.\ Rev.\ Lett.\ {\bf 69}, 1085 
(1992), and references therein.

\bibitem{Hirsch_Fradkin}
J.E. Hirsch and E. Fradkin, Phys.\ Rev. B {\bf 27}, 4302 (1983).

\bibitem{Lebowitz}
G. Benfatto, G. Gallavotti and J. L. Lebowitz, Helv.\ Phys.\ Acta
{\bf 68}, 312 (1995).

\bibitem{Bursill}
R. J. Bursill, unpublished.

\bibitem{White}
S.\ R.\ White, Phys.\ Rev.\ Lett.\ {\bf 69}, 2863 (1992); Phys.\ Rev.\ B 
{\bf 48}, 10 345 (1993);
G. A. Gehring, R. J. Bursill and T. Xiang, Acta Physica Polonica
{\bf 91}, 105 (1997).

\bibitem{Caron}
L. G. Caron and S. Moukouri, Phys.\ Rev.\ Lett.\ {\bf 76}, 4050 (1996);
Phys.\ Rev.\ B {\bf 56}, R8471 (1997).

\bibitem{Jeckelmann1}
E. Jeckelmann and S. R. White, Phys.\ Rev.\ B {\bf 57}, 6376 (1998).
%cond-mat/9710058.

\bibitem{Jeckelmann2}
C. Zhang, E. Jeckelmann and S. R. White, Phys.\ Rev.\ Lett.\
{\bf 80}, 2661 (1998).
%cond-mat/9709187.

\bibitem{notation}
It will be shown (see Fig.\ \ref{softening}) that, in general,
the lowest two neutral excitations are neither purely fermionic
(particle-hole) nor phonon excitations and so we denote them
1- and 2- ``photon'' excitations, as they are the excitations that
would be seen if one- and two- photon absorption experiments
were carried out on a system described by the model.
In the spin language of Nomura and Okamoto (ref.\ \cite{Nomura}), 
$\Delta_1$ and $\Delta_2$ are known as the {\em N\'{e}el} and
{\em dimer} gaps.

%\bibitem{asymptotic}
%There are two possible ways in which
%the charge density wave can form: the charge density
%can increase on either the odd sites or the even sites.
%The ground state and the excited state associated
%with $\Delta_1$ are the symmetric
%and antisymmetric linear combinations of these two states.
%The gap between them rapidly closes  as the lattice size is increased.
%This is analogous to the asymptotic degeneracy of the ground state
%in the Ising phase of the XXZ model, where again the bulk ground
%state is a degenerate doublet.

\bibitem{Affleck}
I. Affleck {\em et al.}, J. Phys.\ A {\bf 22}, 
511 (1989); K. Okamoto and K. Nomura, Phys.\ Lett.\ A {\bf 169}, 433
(1992).

\bibitem{Nomura}
K. Nomura and K. Okamoto, J. Phys.\ A {\bf 27}, 5773 (1994).
%, and references therein.

\bibitem{differences}
This is different from determining the transition point 
in the frustrated Heisenberg model which is spin rotationally 
symmetric ($\Delta_{\text{ch}} \equiv \Delta_1$ for all $N$). For that 
model the crossover point is defined by $\Delta_1 = \Delta_2$.
%: i.e., the crossover of singlet and triplet gaps.

\bibitem{Caron2}
L. G. Caron and C. Bourbonnais, Phys.\ Rev.\ B {\bf 29}, 4230 (1984). 
%The prefactor 8 depends on the choice of cutoff in the derivation. 
%However, this has only a small effect on the predicted phase boundary.

\bibitem{Wu} C. Q. Wu, Q. F. Huang, and X. Sun, Phys.\ Rev.\ B {\bf 52}, 
15683 (1995).

%\bibitem{Zhe} H. Zheng, D. Feinberg, and M. Avignon, Phys.\ Rev.\ B
%{\bf 39}, 9405 (1988).
%
\bibitem{Voit}
J.\ Voit, Rep.\ Prog.\ Phys.\ {\bf 58}, 977 (1995).

\bibitem{Cardy}
J. L. Cardy, J. Phys.\ A {\bf 17}, L385 (1984).

\bibitem{Shankar}
R. Shankar, Int.\ J. Mod.\ Phys.\ {\bf 4}, 2371 (1990).

\bibitem{Cardy1}
J. L. Cardy, J. Phys.\ A %{\bf 19}, L1093 (1986);
{\bf 20}, 5039 (1987).

%%\end{thebibliography}

%%
\end{references}
\end{document}